\documentclass[useAMS,usenatbib,usegraphicx]{mn2e}

\usepackage{times}
\usepackage{amsmath,amsfonts,amssymb}
\usepackage{natbib}
\usepackage{journals}
\usepackage[colorlinks=True,%
            urlcolor=blue,%
            linkcolor=blue,%
            citecolor= blue,%
            pdfauthor={T. Gastine},%
            pdftitle={From solar-like to anti-solar differential rotation in 
cool stars},%
            dvips]{hyperref}

\usepackage[usenames,dvipsnames,svgnames,table]{xcolor}

\title[From solar-like to anti-solar differential rotation in 
cool stars]{From solar-like to anti-solar differential rotation in 
cool stars}
\author[T.~Gastine, R.~K.~Yadav, J.~Morin, A.~Reiners, and J.~Wicht]{T. 
Gastine$^{1}$\thanks{E-mail:
gastine@mps.mpg.de}, R.~K.~Yadav$^{1,2}$, J.~Morin$^{2,3}$, A.~Reiners$^2$ 
and 
J.~Wicht$^1$ \\
$^{1}$Max Planck Institut f\"ur Sonnensystemforschung, Max-Planck Strasse 2., 
37191 Katlenburg-Lindau, Germany\\
$^{2}$ Institut f\"ur Astrophysik, Georg-August-Universit\"at G\"ottingen, 
Friedrich-Hund Platz, 37077 G\"ottingen, Germany\\
$^{3}$  LUPM--UMR5299, CNRS \& Universit\'e Montpellier \textsc{ii}, Place 
E. Bataillon, 34095 Montpellier Cedex 05, France}

\begin{document}

\date{Accepted for publication in MNRAS, 8 November 2013}

\pagerange{\pageref{firstpage}--\pageref{lastpage}} \pubyear{2013}

\maketitle

\label{firstpage}

\begin{abstract}
Stellar differential rotation can be separated into two main regimes: 
solar-like when the equator rotates faster than the poles and anti-solar when 
the polar regions rotate faster than the equator. We investigate the transition 
between these two regimes with 3-D numerical simulations of rotating spherical 
shells. We conduct a systematic parameter study which also includes models from 
different research groups. We find that the direction of the differential 
rotation is governed by the contribution of the Coriolis force in the force 
balance, independently of the model setup (presence of a magnetic field, 
thickness of the convective layer, density stratification). Rapidly-rotating 
cases with a small Rossby number yield solar-like differential rotation, while 
weakly-rotating models sustain anti-solar differential rotation. Close to the 
transition, the two kinds of differential rotation are two possible bistable 
states. This study provides theoretical support for the existence of anti-solar 
differential rotation in cool stars with large Rossby numbers.
\end{abstract}


\section{Introduction}

The solar surface rotates differentially with the equatorial regions 
rotating faster than the poles. In addition, helioseismic measurements revealed 
the internal rotation profile of the Sun: (\textit{i}) the outer 
convective region exhibits significant latitudinal gradients of shear; 
(\textit{ii}) a strong radial differential rotation is observed at the bottom 
of the convective zone forming the tachocline; (\textit{iii}) and the radiative 
core rotates nearly uniformly \citep[e.g.][]{Thompson03}.

In cool stars other than the Sun, the surface differential rotation can be 
inferred from different measurements techniques encompassing Doppler imaging 
\citep[e.g.][]{Collier02}, Fourier transform of the spectral lines 
\citep[e.g.][]{Reiners02} or period detection in the photometric measurements
\citep[e.g.][]{Reinhold13}. The latitudinal differential rotation in stars is 
usually described by a single-parameter law of the form 
$\Omega(\theta)=\Omega_e(1-\alpha \sin^2 \theta$), $\theta$ being the latitude 
and $\Omega_e$ the angular velocity at the equator. Differential 
rotation is then usually categorised as ``solar-like'' when $\alpha > 0$ 
($\alpha_\odot = 0.2$), or as ``anti-solar'' when the polar regions rotate 
faster than the equator (i.e. $\alpha < 0$). For main and pre-main sequence 
stars, observations of absolute surface shear show some dependence on rotation 
period and effective temperature \citep{Barnes05,AvER12,Reinhold13}. 
Information on the sign of differential rotation is very sparse because 
observational signatures are very subtle (Fourier technique) or only the 
absolute value is obtained (photometric technique). Up to now, anti-solar 
differential rotation has only been reported for a handful of K giant stars 
observed with the Doppler imaging technique 
\citep[e.g.][]{Strassmeier03,Weber05,Kovari07}. We may thus wonder what 
determines the sign of differential rotation in cool stars?

The first theoretical approach to model stellar differential rotation relies on 
hydrodynamical mean-field models \citep[e.g.][]{Rudiger89}.  In a similar way 
as in the mean-field dynamo models, the velocity components $u_i$ are therefore 
decomposed into a mean-field contribution $\bar{u}_i$ and a fluctuating part 
$u_i'$. The quadratic correlations of the fluctuating quantities, such as 
Reynolds stresses ${\cal Q}_{ij} = \overline{u_i' u_j'}$, are then parametrised 
as functions of the mean-field quantities only. Reynolds stresses are for 
instance expanded assuming ${\cal Q}_{ij}=\Lambda_{ijk}\bar{\Omega}_k - 
N_{ijkl}\partial \bar{u}_k / \partial x_l$, where $\Lambda_{ijk}$ and 
$N_{ijkl}$ are third and fourth order tensors, respectively. The 
parametrisation of the velocity correlations ${\cal Q}_{ij}$ thus involves some 
free coefficients (turbulent viscosity for instance) that need to be set to 
ensure the closure of the mean-field model. Despite these approximations, 
mean-field approaches were quite successful in predicting a weak dependence of 
the surface shear on the rotation rate and a strong correlation with the 
effective temperature as observed on the main sequence stars \citep{Kuker11}.  
In addition, these models have a strong prediction concerning the sense of the 
differential rotation and predominantly yield solar-like $\Omega(\theta)$ 
profiles \citep{Kitchatinov99}. Anti-solar differential rotation can only be 
maintained in case of very strong meridional circulation \citep{Kitchatinov04}.

Alternatively, stellar differential rotation can be modelled using 3-D 
hydrodynamical and dynamo models of rotating convection in spherical geometry. 
In that case, the differential rotation is maintained by the interaction of 
turbulent convection with rotation. Despite their own limitations (high 
diffusivities and moderate density contrasts), 3-D models 
allow to fully take into account the 
nonlinearities involved in the angular momentum transport. In contrast with 
mean-field approaches, no parametrisation of Reynolds stresses 
is required in 3-D simulations. Although a large number of such simulations 
yield solar-like differential rotation, they have also frequently produced 
anti-solar differential rotation over a broad range of parameters and model 
setups \citep[e.g.][]{Gilman77,Glatz5,Aurnou07,Steffen07,Matt11, 
Kapyla11a,Bessolaz11,Gastine13a}. The differential rotation direction is 
suspected to be controlled by the relative contribution of buoyancy and 
Coriolis force in the global force balance \citep{Gilman77,Aurnou07}. 
Cases where rotation dominates the force balance yield prograde equatorial 
azimuthal flows, while a weak rotational influence leads to anti-solar 
differential rotation. As shown in previous parameter studies, these two 
regimes can be well separated by a critical convective Rossby number of 
unity \citep{Gilman77}, independently of the background density stratification 
\citep[][hereafter GW12, GWA13]{Gastine12,Gastine13a}.

The present work extends these studies to a broader range of parameters to 
investigate the zonal flow transition in 3-D models in a systematic way. For 
the sake of generality, we also incorporate data of different research groups 
who reported anti-solar differential rotation in their models.

\section{Hydrodynamical model}


We consider numerical simulations of an anelastic ideal gas in spherical shells 
rotating at a constant rotation rate $\Omega_0$. A fixed entropy contrast 
$\Delta s$ between the inner and the outer boundary drives the convective 
motions. Our numerical models are computed using the anelastic spectral code 
MagIC \citep[][GW12]{Wicht02} that has been validated against hydrodynamical 
and dynamo benchmarks \citep{Jones11}. We non-dimensionalise the MHD equations 
using $\Omega_0^{-1}$ as the time unit and the shell thickness $d=r_o-r_i$ as 
the reference lengthscale. The anelastic system of equations is then governed 
by four dimensionless parameters

\begin{equation}
 E=\dfrac{\nu}{\Omega_0 d^2},\,Ra=\dfrac{g_o d^3 \Delta s}{c_p \nu \kappa},\, 
Pr=\dfrac{\nu}{\kappa},\,Pm=\dfrac{\nu}{\lambda},
\end{equation}
where $\nu$, $\kappa$, and $\lambda$ are the constant kinematic, thermal and 
magnetic diffusivities and $g_o$ is the gravity at the outer boundary. Details 
of the numerical implementation are extensively discussed by \cite{Jones11} and 
GW12. Differential rotation maintained in 3-D models is suspected to be 
sensitive  to the relative contribution of buoyancy and Coriolis force in the 
force balance. The ratio between these two forces can be roughly 
assessed by the so-called \emph{convective Rossby  number}, defined by 
$Ro_c=\sqrt{Ra\,E^2/Pr}$.

The converged solution of a numerical simulation is then characterised by 
several diagnostic parameters. The rms flow velocity is 
given in units of the Rossby number $Ro'=u'_\text{rms}/\Omega_0 d$, 
where primed quantities correspond to the non-axisymmetric contribution.
The typical flow lengthscale $\ell$ is defined as $\ell = \pi d/\bar{l}_u$, 
where $\bar{l}_u$ is the mean spherical harmonic degree obtained from the 
kinetic energy spectrum \citep[e.g.][]{Christensen06,Schrinner12}. A \emph{local 
Rossby number} $Ro_\ell = u'_\text{rms}/\Omega_0 \ell$ can then be used to 
evaluate the relative contribution of inertia and Coriolis force to the global 
force balance. Differential rotation is quantified by the amplitude of the 
equatorial surface zonal flow:

\begin{equation}
 \alpha_e = \dfrac{\bar{u}_\phi(r=r_o,\theta=0)}{\Omega_o r_o} = 
\dfrac{d\Omega(r=r_o,\theta=0)}{\Omega_o},
\end{equation}
where overbars denote axisymmetric quantities.


\begin{figure}
 \centering
 \includegraphics[width=8.4cm]{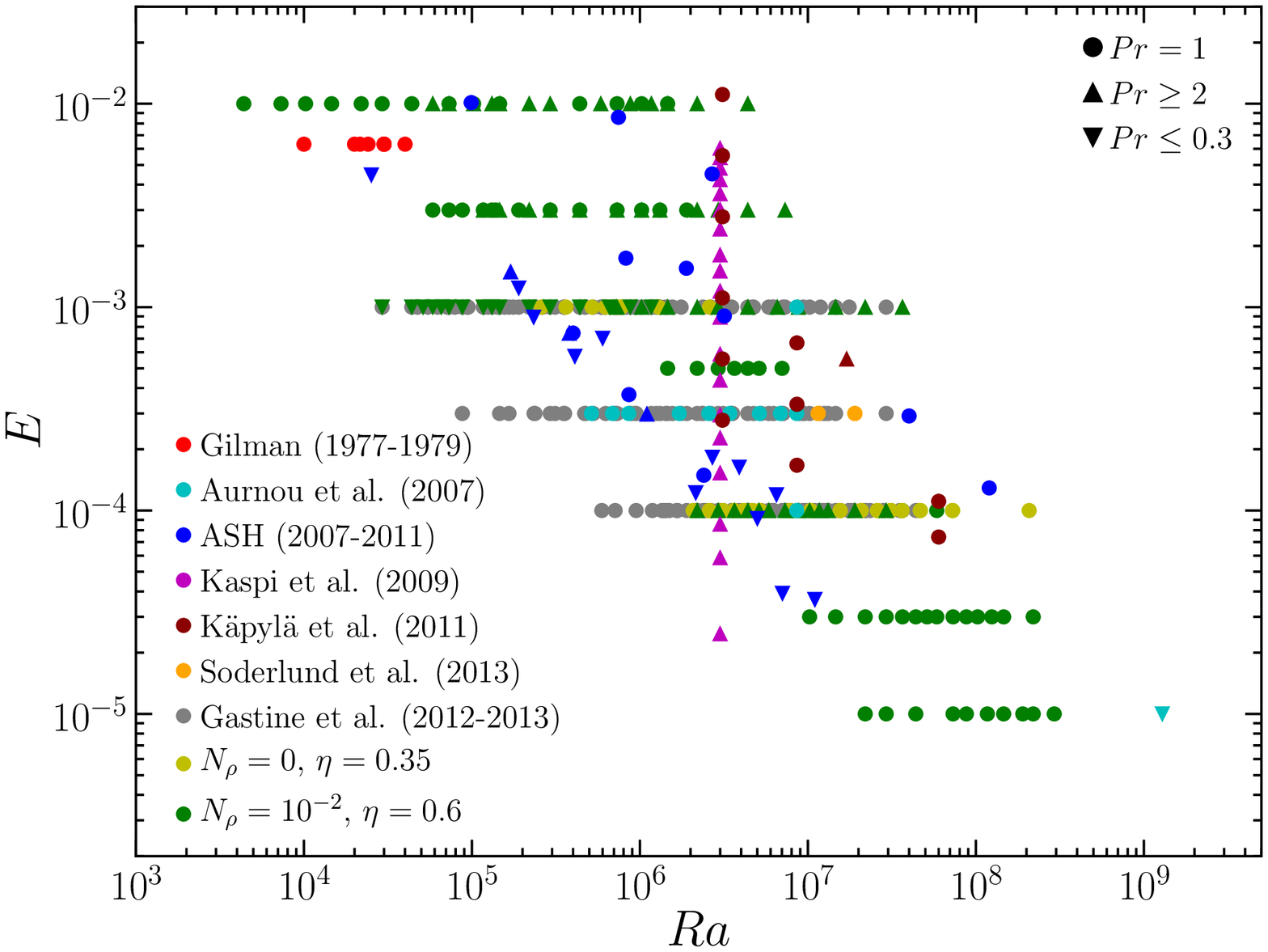}
 \caption{Dimensionless control parameters explored by various numerical models 
computed with different codes. Data have been gathered and adapted from
\citet{Gilman77,Gilman79a}; \citet{Aurnou07}, ASH
\citep{Ballot07,Browning08,Brun09,Brown09,Matt11,Bessolaz11}; \citet{Kaspi09};
\citet{Kapyla11a}; \citet{Soderlund13}; GW12 and GWA13.}
 \label{fig:raEk}
\end{figure}

Our previous parameter studies were dedicated to the effects of the density 
stratification on the differential rotation (GW12, GWA13). They 
assumed $Pr=1$  and covered a limited range of Ekman numbers
($E=10^{-3}-10^{-4}$). To extend the coverage of the parameter space, we have 
computed here 150 new cases which span the range of $10^{-5}<E<10^{-2}$, 
$10^3<Ra<5\times10^8$ and $Pr\in[0.1, 1, 10]$. We consider here 
non-magnetic nearly Boussinesq models (i.e. $N_\rho = 
\ln(\rho_{\text{bot}}/\rho_{\text{top}})=10^{-2}$) in a thin spherical shell of 
aspect ratio $\eta=r_i/r_o=0.6$. To investigate how the magnetic field 
influences differential rotation, we also consider a few Boussinesq dynamo 
models with $\eta=0.35$ and $Pm=1$.
 
Furthermore, we include additional data from published studies which 
encompasses Boussinesq \citep[e.g.][]{Aurnou07}, anelastic
\citep[e.g.][ASH]{Gilman77} and fully compressible 3-D models \citep{Kapyla11a}.
To our knowledge, all the data reporting anti-solar differential rotation have 
been gathered in Fig.~\ref{fig:raEk}, provided control and relevant diagnostic 
parameters (i.e. $\alpha_e$) were accessible. Note that to ease the comparison 
between the different setups, the Rayleigh numbers have been rescaled in 
Fig.~\ref{fig:raEk} to use the entropy gradient at mid-depth, i.e.  $Ra = g_o 
d^3 |ds/dr|_m / c_p \nu \kappa$. This provides a better way of
comparing different reference state models \citep[see][GWA13]{Kaspi09}.

\section{Results}

\subsection{Differential rotation regimes}

\begin{figure}
 \centering
 \includegraphics[width=8.4cm]{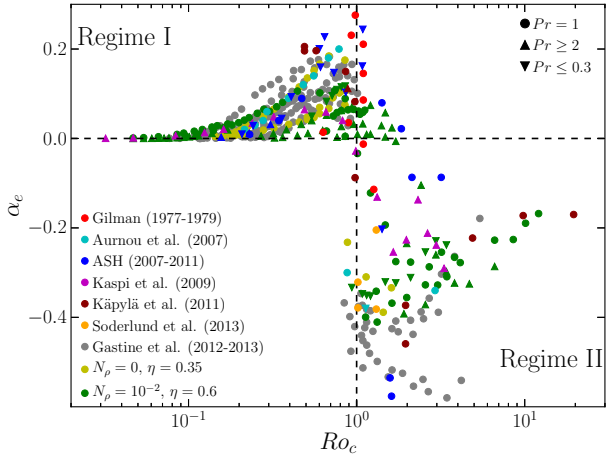}
 \caption{Amplitude of the surface zonal flows at the equator in units of 
$\alpha_e=\bar{u}_\phi/\Omega_0 r_o$ as a function of $Ro_c$ for 
the numerical models of Fig.~\ref{fig:raEk}.}
 \label{fig:roEk}
\end{figure}

Figure~\ref{fig:roEk} shows the surface differential rotation amplitude 
$\alpha_e$ as a function of $Ro_c$ for the Fig.~\ref{fig:raEk} dataset. When 
Coriolis forces dominate the force 
balance (i.e. $Ro_c  \ll 1$, regime I), the equatorial zonal flow is prograde 
and its amplitude increases with $Ro_c$. A relatively sharp transition to 
retrograde zonal winds (or anti-solar differential rotation) then occurs close 
to $Ro_c \sim 1$. Although the dataset is scattered, the 
retrograde equatorial flow is on average stronger than in regime I and 
reaches values of $\alpha_e \sim -0.4$ for $Ro_c \sim 1$. When buoyancy starts 
to dominate the force balance (i.e. $Ro_c \gg 1$, regime II), the 
differential rotation decreases suggesting a possible third regime 
where turbulent motion gradually suppress the mean flows \citep[$Ro_c > 10$, 
see GWA13 and][]{Brummell98}.

Despite differences in size of the convective layer, values of the control 
parameters, definition of the Rayleigh number, choice of thermal boundary 
conditions and so on, the transition between regimes I and II is well captured 
by $Ro_c$, with all the data points concentrating in the top-left and 
bottom-right quadrants.

\begin{figure}
 \centering
 \includegraphics[width=8.4cm]{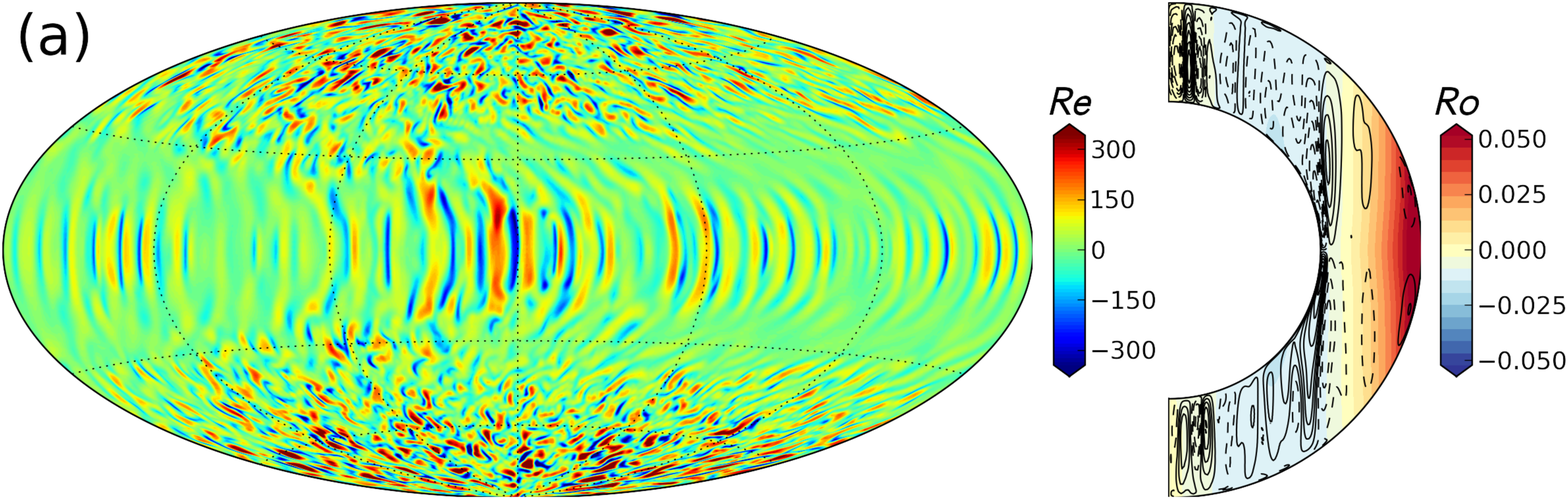}
 \includegraphics[width=8.4cm]{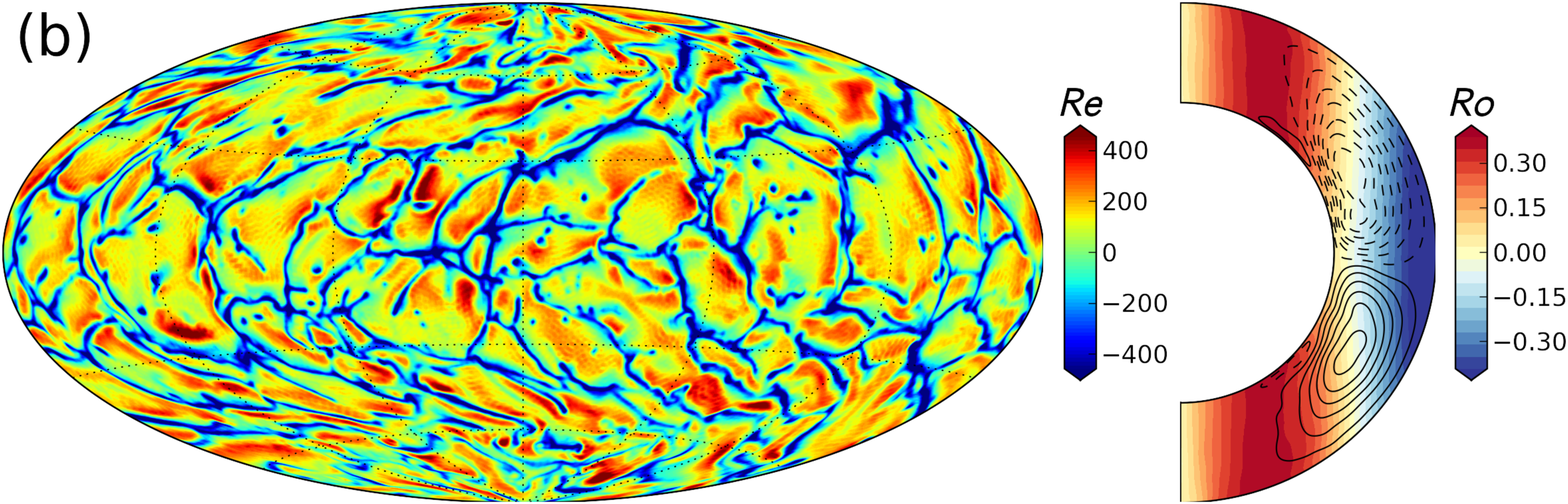}
 \caption{Left panels: radial velocity at $r=0.95\,r_o$. Right panels: 
time-averaged zonal velocity (colored levels) and meridional circulation 
(solid and dashed lines). Case (\textbf{a}) corresponds to $N_\rho=3$, 
$\eta=0.6$, $E=10^{-4}$, $Ra=9\times 10^{6}$, $Pr=1$, case (\textbf{b}) to
$N_\rho=3$, $\eta=0.6$, $E=10^{-3}$, $Ra=4\times 
10^{6}$, $Pr=1$. Radial velocity is given in Reynolds number units ($u_r 
d/\nu$) while zonal flows are expressed in Rossby number units.}
  \label{fig:ex}
\end{figure}

To illustrate the differences in the differential rotation patterns in the two 
regimes, Fig.~\ref{fig:ex} shows radial velocity and zonal flows for two 
selected models. In the rotation-dominated regime ($Ro_c=0.09$, upper panels) 
convective columns aligned with the rotation axis are visible at low 
latitudes.  They are accompanied at higher latitudes by small-scale
time-dependent convective cells. Due to the curvature of the spherical shell, 
the convective columns are slightly tilted in the prograde direction and give 
rise to Reynolds stresses \citep[a statistical correlation between the 
convective flow components, see][]{Busse83,Christensen02}. Reynolds stresses 
maintain a positive flux of angular momentum away from the rotation axis which 
is responsible for the observed differential rotation. The pair of geostrophic 
zonal flows with an eastward equator and westward poles is typical in this 
regime \citep[e.g.][GW12]{Kapyla11a}. In contrast, when buoyancy becomes a 
first-order contribution in the force balance ($Ro_c=4$, lower panels), the 
convective features lose their preferred alignment with the rotation axis 
and the zonal flow direction reverses.  The equatorial jet becomes retrograde 
and is flanked by two prograde zonal winds inside the tangent cylinder. The 
anti-solar differential rotation observed here can be attributed to the mixing 
of angular momentum by the turbulent convective motions 
\citep[e.g.][]{Gilman79,Aurnou07}. As demonstrated by GWA13, the angular 
momentum per unit mass ${\cal M}$ is thus a conserved quantity such that

\begin{equation}
 {\cal M} = \bar{u}_\phi s + \Omega_0 s^2 = \text{const.} = \zeta(\eta, 
N_\rho)  \Omega_0 r_o^2,
\end{equation}
where $s$ is the cylindrical radius and $0<\zeta(\eta, N_\rho) < 1$ 
depends on the background density stratification, the size of the 
convective zone and the efficiency of the angular momentum mixing.
Using $\Omega_0 r_o^2$ to non-dimensionalise this equation leads to the 
following formulation of the differential rotation in regime II:

\begin{equation}
  Ro = \dfrac{\bar{u}_\phi}{\Omega_0 r_o } = \zeta(\eta, N_\rho) 
\dfrac{r_o}{s}-\dfrac{s}{r_o}.
\label{eq:zfprof}
\end{equation}
Comparisons between the zonal flow profiles and this theoretical prediction 
give a good agreement for models with $Ro_c \gtrsim 1$
\citep[][GWA13]{Aurnou07}.

Meridional circulation patterns change when differential rotation changes sign
\citep[e.g.][]{Matt11,Bessolaz11}. In the upper panel of Fig.~\ref{fig:ex}, 
multiple small-scale meridional circulation cells are observed, while the second 
model shows only one large-scale cell in each hemisphere. This transition 
results from a change in the spatial variations of the azimuthal force balance 
between viscous and Reynolds stresses \citep[a mechanism sometimes known as 
``gyroscopic pumping'', e.g.][]{Miesch11}.

\begin{figure}
 \centering
 \includegraphics[width=8.4cm]{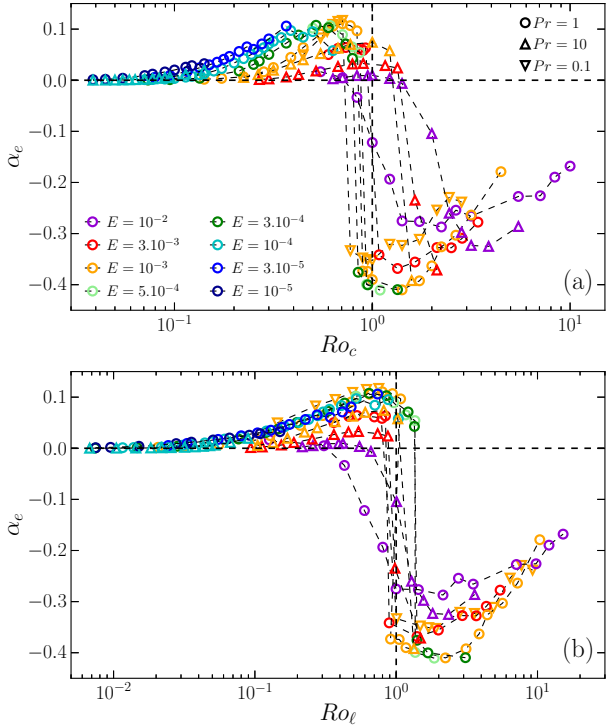}
 \caption{(\textbf{a}) $\alpha_e$ as a function of 
$Ro_c$ and (\textbf{b}) as a function of $Ro_\ell$. All the numerical models 
have $N_\rho=10^{-2}$ and 
$\eta=0.6$.}
 \label{fig:roEqNrho0}
\end{figure}

Figure~\ref{fig:roEqNrho0}a shows the same quantities as Fig.~\ref{fig:roEk} 
for a consistent subset of nearly Boussinesq numerical models ($N_\rho=10^{-2}$) 
with  $\eta=0.6$. This subset is partly composed by the Boussinesq models of 
GW12 and GWA13 and partly by the additional cases computed for the present 
study. Considering the same reference model for the whole subset allows to more 
accurately scrutinise the zonal flow transition. While the regime change occurs 
in the range $0.8<Ro_c<2$, some parameter dependence is still noticeable. For 
instance, the transition is rather gradual for large Ekman numbers 
($E=10^{-2}$, magenta symbols) and becomes sharper when the Ekman number is 
lowered. Moreover, the Prandtl number dependence does not seem to be perfectly 
captured by $Ro_c$. In fact, the zonal flow transition in the numerical models 
with $Pr=10$ ($Pr=0.1$) takes place at higher (lower) values of $Ro_c$ than the 
$Pr=1$ cases. As our dataset is limited to relatively large Ekman numbers for 
$Pr\neq 1$, we might however speculate that such $Pr$ dependence vanishes at 
low Ekman numbers. As shown in Fig.~\ref{fig:roEqNrho0}b, the zonal flow 
transition is better captured when $\alpha_e$ is plotted against the local 
Rossby number $Ro_\ell$. This reduction of the dispersion is expected as $Ro_c$ 
is only a rough proxy of the convective Rossby number, while $Ro_\ell$ is a 
measure of the actual local Rossby number of a numerical model. A precise 
estimate of $Ro_\ell$ for the whole dataset of models shown in 
Fig.~\ref{fig:roEk} would thus also help to reduce the observed dispersion.

\subsection{Zonal flow bistability}

\begin{figure}
 \centering
 \includegraphics[width=8.4cm]{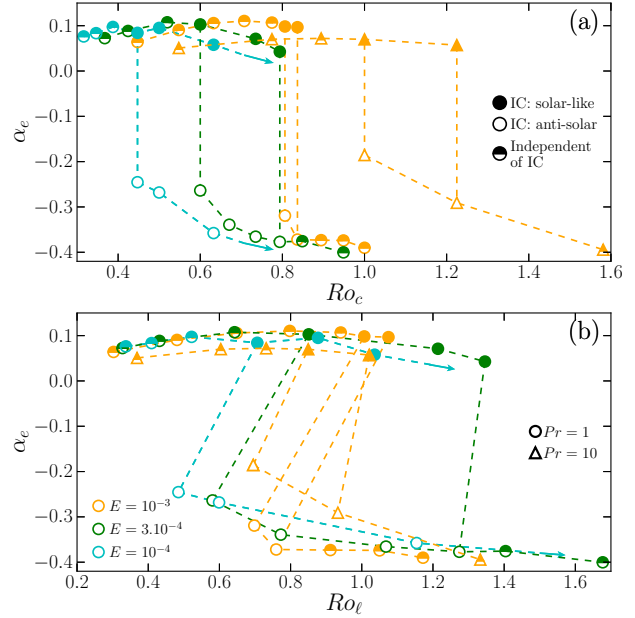}
 \caption{(\textbf{a}) $\alpha_e$ as a function of $Ro_c$   and (\textbf{b}) as 
a function of $Ro_\ell$. Selection of numerical models 
with $N_\rho=10^{-2}$ and $\eta=0.6$ to illustrate the bistability of the 
zonal flow. The dependence on the initial conditions (IC) is shown by 
different symbol fillstyles. The blue arrows indicate possible continuation of 
the hysteresis 	loop for the $E=10^{-4}$ cases.}
 \label{fig:bistab}
\end{figure}

We find several cases of bistability where the two kinds of differential 
rotation are stable at identical parameters (i.e. $Ra$, $E$ and $Pr$) when 
$Ro_c\sim Ro_\ell \sim 1$. The initial condition then 
selects which differential rotation profile will be 
adopted by the converged solution. As shown on Fig.~\ref{fig:bistab},
starting from a model with a solar-like differential rotation and 
increasing $Ro_c$ (or $Ro_\ell$) maintains a solution with the same kind of 
differential rotation for $0.5<Ro_\ell<1.3$ before falling on the other 
branch at higher $Ro_\ell$. Alternatively, if one initiates this model 
with $\alpha_e < 0$ and decreases $Ro_c$, the solution may remain on that 
branch. Once again, the $Pr$ dependence on the bistability region seems 
to be better captured when one considers $Ro_\ell$ instead of $Ro_c$. The 
hysteresis loop is relatively narrow for $E=10^{-3}$ and becomes wider at 
$E=3\times 10^{-4}$.  At $E=10^{-4}$, it becomes numerically too demanding to 
further investigate the extent of the two branches. Hence, an 
Ekman number dependence cannot be completely ruled out and the 
extent of the bistability region might increase further when $E$ is lowered.

\subsection{Magnetic field influence}

\begin{figure}
 \centering
 \includegraphics[width=8.4cm]{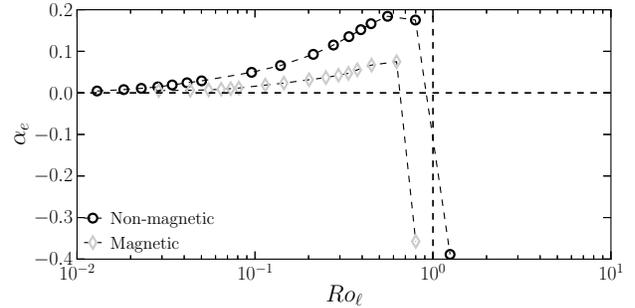}
 \caption{$\alpha_e$ as a function of $Ro_\ell$  for non-magnetic and 
magnetic ($Pm=1$) models with $N_\rho=0$, $\eta=0.35$, $E=10^{-4}$ and $Pr=1$.}
\label{fig:mag}
\end{figure}

To investigate if the zonal flow transition is affected by the presence of
magnetic field, we compute two sets of Boussinesq models with 
$\eta=0.35$: one consists of non-magnetic cases, while the other 
contains their dynamo counterparts. Figure~\ref{fig:mag} shows that in both 
cases the transition between regimes I and II occurs around $Ro_\ell 
\sim 1$. Due to the influence of the magnetic 
field on the convective flow velocity and lengthscale, the exact value 
of $Ro_\ell$ at the transition is slightly lower in the dynamo models. In 
the rotation-dominated regime, the magnetic cases have significantly weaker 
zonal flows than the non-magnetic ones. In contrast, hydrodynamical and dynamo 
models yield similar zonal flow amplitude in regime II, confirming the previous 
findings by \cite{Soderlund13}. These differences can be attributed to the 
relative efficiency of the magnetic braking. The quenching of the differential 
rotation by Lorentz forces is indeed more pronounced when the magnetic field 
has a significant large-scale contribution  \citep[$Ro_\ell < 1$, 
][]{Yadav13}.

\section{Discussion}

We investigate the transition between solar-like and anti-solar differential 
rotation in rotating spherical shells. We extend previous studies 
\citep{Gastine12,Gastine13a} with a new set of models which covers a broader 
range of control parameters. We also include models published by various groups 
in our analysis.

From this set of simulations we confirm previous findings that the 
direction of differential rotation is determined by the value of the 
convective Rossby number defined as $Ro_c=\sqrt{Ra\,E^2/Pr}$. 
In the rotation-dominated regime (regime I, $Ro_c < 1$), the 
differential rotation 
is solar-like, i.e. the equator rotates faster than the poles. When buoyancy 
dominates the force balance ($Ro_c > 1$), the turbulent convective 
motions homogenise the angular momentum, which leads to anti-solar differential 
rotation profiles. The regime transition takes place at $Ro_c \sim 
1$,  independently of the details of the model (density stratification, 
thickness of the convective layer and so on). We show that the local Rossby 
number $Ro_\ell$ -- a good proxy of the relative contribution of Coriolis force 
and inertia in the force balance \cite[][]{Christensen06} -- helps to  better 
separate the two regimes. Close to the transition ($0.5 < Ro_\ell < 1.5$), the 
two kinds of differential rotation are two possible stable states at the same 
parameter values, forming a bistable region. The presence of a magnetic field 
reduces the amplitude of differential rotation in regime I without affecting the 
regime change at $Ro_\ell \sim 1$. 

It should be however noted that global numerical models always operate in 
a parameter regime far from the stellar values due to their large diffusivities 
(i.e. small Rayleigh and large Ekman numbers). Hence, the existence of 
additional dynamical regimes cannot be ruled out at realistic parameters.
Nonetheless, anti-solar differential rotation is systematically found in 
weakly-rotating 3-D simulations in contrast with the mean-field results.
A closer comparison between mean-field predictions and 3-D 
simulations is therefore desirable to better establish the limits of validity of 
such mean-field approaches \citep[e.g.][]{Kapyla11a}.

Our results provide theoretical support for the 
existence of slowly rotating cool stars exhibiting anti-solar differential 
rotation. A further validation of our prediction requires to estimate $Ro_\ell$ 
in stellar convective zones. We adopt $Ro_\text{emp}$, the ratio of the 
rotation period $P_\text{rot}$ and the turnover time of convection 
$\tau_\text{conv}$, as our best available proxy for $Ro_\ell$ 
\citep[e.g.][]{Gastine13}. With 
$P_\text{rot}=25$~d  and $\tau_\text{conv} = 12-50$~d \cite[][]{Reiners10},
the solar Rossby number lies in the range $0.5 <Ro_\text{emp} 
< 2$. This suggests that the Sun might be at the limit of the 
rotation-dominated regime and that stars with Rossby number just above the solar 
value could exhibit strong anti-solar differential rotation. Claims of 
anti-solar differential rotation are so far restricted to K giants. Most of 
these stars  are in binary systems where tidal effects likely have an impact on 
the surface shear \citep[e.g.][]{Kovari07}. The K giant HD~31993 is the only 
single giant for which a significant anti-solar differential rotation is 
reported \citep[$\alpha=-0.125$,][]{Strassmeier03}. For this star 
$P_\text{rot}=25.3$~d and $\tau_\text{conv}\simeq 25$~d \citep{Gunn98} 
yield $Ro_\text{emp} \simeq 1$, a value close to the threshold but compatible 
with $\alpha < 0$. 

Measuring differential rotation for stars clearly in the $Ro > 1$ regime 
remains challenging. Doppler imaging or line profile analysis are sensitive to 
the sign of differential rotation but suffer from some limitations. Doppler 
imaging indeed relies on the 
presence of large spots at the stellar photosphere which is not expected for 
$Ro_\text{emp}>1$. Line profile analysis requires a minimum  
rotational velocity $v\sin{i}_\text{min}\sim 10-20$~km.s$^{-1}$. This is 
incompatible with $Ro_\text{emp}>1$ for cool main sequence stars. Although 
\cite{AvER12} observed line profile shapes attributable to $\alpha \lesssim 0$ 
for dwarf stars with $Ro_\text{emp} < 1$, they attributed these signatures to 
the presence of cool polar spots. With their larger radii, weakly active 
evolved giant stars might be more suitable targets. Space missions CoRoT and 
Kepler collect high-precision photometric data for a vast sample of stars. 
Although this technique cannot directly determine the sign of $\alpha$, a 
regime change in the differential rotation might still be captured. Our 
numerical models indeed suggest a relatively sharp rise in $|\alpha|$ at the 
transition between solar and anti-solar differential rotation. Latest results 
based on moderate to fast rotators ($P_\text{rot} < 45$~d) by \cite{Reinhold13} 
suggest a possible increase of $|\alpha|$ with the Rossby number stressing the 
need for further analysis of slowly-rotating Kepler stars.

\vspace{-0.6cm}

\section*{Acknowledgements}
{\small
We thank P.~K\"apyl\"a for providing us the parameters of his numerical 
models and A.~S.~Brun for fruitful discussion. Computations have been carried 
out on the GWDG computer facilities in G\"ottingen and on HRLN in Hannover. We 
acknowledge funding from the Deutsche Forschungsgemeinschaft (DFG) through 
Project SFB 963/A17 and through the Special Priority Program 1488.}

\vspace{-0.5cm}

\bibliographystyle{mn2e}

{\footnotesize
\bibliography{biblio}}

\bsp

\label{lastpage}

\end{document}